\documentclass[reprint,titlepage,amsmath,amssymb,aps,physrev,floatfix,superscriptaddress]{revtex4-2}

\usepackage{graphicx}
\usepackage[utf8]{inputenc}
\usepackage[T1]{fontenc}
\usepackage{newtxtext}
\usepackage{newtxmath}
\usepackage{microtype} 
\usepackage{physics} 
\usepackage{siunitx}
\usepackage[svgnames]{xcolor} 
    \definecolor{accent}{HTML}{df2d16}
\usepackage{tikz}
    \usetikzlibrary{arrows.meta, calc, decorations.pathmorphing, decorations.markings}
\usepackage{pgfplots} 
    \pgfplotsset{compat=1.18}
    \usepgfplotslibrary{groupplots}
\usepackage{xr} 
\usepackage[allcolors=accent,colorlinks,pdfusetitle,pdfauthor={Níckolas de Aguiar Alves and Bruno Arderucio Costa}]{hyperref} 
\usepackage{orcidlink} 

\begin{document}
\title{Liénard--Wiechert potentials and the electromagnetic memory effect}
\thanks{The following article has been accepted by the \href{https://pubs.aip.org/aapt/ajp}{\emph{American Journal of Physics}}. After it is published, it will be found at \url{https://doi.org/10.1119/5.0319859}.}

\author{\firstname{Níckolas} de \surname{Aguiar Alves}\,\orcidlink{0000-0002-0309-735X}}
\email{alves.nickolas@ufabc.edu.br}
\affiliation{Center for Natural and Human Sciences, \href{https://ror.org/028kg9j04}{Federal University of ABC}, Avenida dos Estados 5001, Bangú, Santo André, São Paulo 09280-560, Brazil}

\author{Bruno \surname{Arderucio Costa}\,\orcidlink{0000-0001-5182-2010}}
\email{bcosta@troy.edu}
\affiliation{Center for Relativity and Cosmology, \href{https://ror.org/029jj9438}{Troy University}, Troy, Alabama 36082, USA}

\begin{abstract}
Classical electrodynamics is one of the most well-tested and understood theories in physics. After more than a century of history, it may be surprising that such an established theory still makes new predictions that have not yet been experimentally verified. A noteworthy example is the memory effect---a prediction that an electromagnetic wave can leave a lasting influence long after it has passed. This influence is manifested in a velocity ``kick'' on a test charge. This simple remark lies at the heart of modern investigations of the low-energy behavior of gravity, electrodynamics, and gauge theories and awaits confirmation (or refutation) through experiments. From a pedagogical perspective, this current research topic beautifully epitomizes how standard concepts from undergraduate electrodynamics can still lead to new physics. In this work, we use the Liénard--Wiechert solutions for the electromagnetic fields of moving charges to understand what the memory effect is, where it comes from, and how it could be experimentally probed in the near future. We also discuss the connections between the memory effect and other important topics in fundamental physics, as well as the search for memory in modern gravitational wave observatories.
\end{abstract}

\maketitle

\section{Introduction}\label{sec: introduction}
    Waves surround us. We listen to them when we pay attention, with our eyes closed, to the birds chirping. We see them when we look up, in perfect silence, at the stars. We now notice how they impart the most subtle diffraction patterns in an interferometer when two black holes merge in a galaxy far away. 

    While the simplest waves are continuous, with plane waves as a leading example, we usually experience pulses. A loud clap disturbs the air. The disturbance travels at the speed of sound until it reaches our ears. We hear the clap, and the sound is gone. The perturbation continues to propagate past us, and we will never detect it again. Or so it seems. 

\begin{figure}[tb] 
        \centering
        \includegraphics{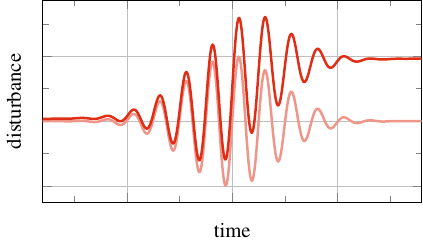}
        \caption{Sketch of a pulse with memory (front, dark line). After the passage of the pulse, the excited field does not return to its original value. Compare to the pulse in the background (light line), which returns to the original value.}
        \label{fig: memory-pulse}
    \end{figure}

    Although we often think of pulses as ephemeral, the wave equation in four dimensions allows them to leave a permanent imprint. The disturbance does not need to vanish after the pulse is gone; it can settle to a constant value. In this case, we say the pulse has memory, and an example is pictured in Fig. \ref{fig: memory-pulse}. The system ``remembers'' the passage of a wave pulse and is permanently altered by it. This memory manifests in different ways in different systems.

    This phenomenon was first predicted for gravitational waves in general relativity, in a paper by \textcite{zeldovich1974RadiationGravitationalWaves}. They noticed that the physical distance between two freely falling particles could be permanently shifted by the passage of a gravitational wave. In this sense, spacetime remembers the gravitational wave through the permanent shift in the distance between nearby objects. Much later, \textcite{bieri2013ElectromagneticAnalogGravitational} noticed that a similar ``memory effect'' is also predicted in classical electrodynamics. After a charged particle interacts with an electromagnetic wave, it may be left with a finite change in its velocity, thus remembering the passage of the pulse.
    
    To see this, let us consider a test charge \(q\) with mass \(m\). We assume an electromagnetic wave passes by. When it does, it interacts with the charge and affects its motion. The acceleration of \(q\) due to the passage of the electromagnetic wave is ruled by the Lorentz force law,
    \begin{equation}\label{eq: lorentz-force-law}
        \vb{F} = q (\vb{E} + \vb{v} \cp \vb{B}).
    \end{equation}
    If the test charge moves slowly, the magnetic force can be neglected. In this case, we may approximate the acceleration of the charge by
    \begin{equation}\label{eq: acceleration-test-charge-for-kick}
        \vb{a} = \frac{q}{m} \vb{E}.
    \end{equation}
    Upon integrating this equation, we find
    \begin{equation}\label{eq: kick-general-formula}
        \Delta \vb{v} = \frac{q}{m} \int \vb{E} \dd{t},
    \end{equation}
    where the integral runs from early to late times. In other words, the integral starts before the passage of the wave and ends after it. This expression depends only on the approximation that the magnetic force is negligible, which amounts to assuming the test charge moves slowly.

    Equation \eqref{eq: kick-general-formula} is deceptively simple. It states at first that the final velocity of the particle does not need to coincide with its initial velocity, and the difference is encoded in the time integral of the electric field. This particular change has, however, some very interesting properties. For example, two comments are already in place \cite{bieri2024ExperimentMeasureElectromagnetic}.
    \begin{enumerate}
        \item The change in the velocity is linear in the electromagnetic field. This contrasts with the electromagnetic momentum density, which is \emph{quadratic} in the fields. The standard expression for radiation pressure is also quadratic (see, for example, Ref. \onlinecite{griffiths2023IntroductionElectrodynamics}, Sec. 9.2.3). Equation \eqref{eq: kick-general-formula} must then capture a different effect.
        \item Equation \eqref{eq: kick-general-formula} predicts a change in velocity along the transverse direction of the electromagnetic wave. Radiation pressure, on the other hand, points in the longitudinal direction---see, as before, Ref. \onlinecite{griffiths2023IntroductionElectrodynamics}, Sec. 9.2.3.
    \end{enumerate}
    In other words, the prediction of Eq. \eqref{eq: kick-general-formula} is not the usual way in which we imagine radiation interacting with charges. In fact, these properties resemble a different scenario: electrostatics. The change in the momentum of a test particle due to the Coulomb field over a certain time interval is precisely given by Eq. \eqref{eq: kick-general-formula}. Nevertheless, the Coulomb field and radiation behave very differently. For example, the Coulomb field falls off as \(1/r^2\) at large distances, while radiation falls off as \(1/r\). Surprisingly, it is verified that Eq. \eqref{eq: kick-general-formula} can have a nonzero \(\order{1/r}\) term. This velocity ``kick'' on a test charge is known as the electromagnetic memory effect. It is hence an intrinsically radiative phenomenon that leaves a lasting effect on test charges. While not yet observed experimentally, the memory of an electromagnetic wave may live on in the velocities of the charges it interacted with. 

    The memory effect is interesting in its own right. It is an example of a prediction of Maxwell electrodynamics that has not been experimentally verified yet. However, it has attracted much attention due to its connections to other topics in physics. This effect should be seen at large distances from the source, in the sense that \(r\) is large enough for us to neglect the \(1/r^2\) term that is due to the Coulomb field. This means memory is a property of the low-frequency regime of electrodynamics. As we will discuss, it has recently been noticed that this regime is extremely rich---not only in electrodynamics, but in other theories as well. There is hope that understanding it could be a path to better comprehending quantum gravity, and these investigations have unraveled connections between the memory effect, new symmetries that appear at very large distances, and the behavior of particles with very low energy in quantum field theory. These new symmetries are known as ``large gauge transformations,'' while the behavior of very-low-energy particles is encoded in the so-called ``soft theorems.''

    Our goal in this text is to explain, in terms of a few calculations accessible to an undergraduate electrodynamics course, how the memory effect is an observable prediction of Maxwell's theory. More concretely, we derive a special case of the electromagnetic memory effect first obtained by \textcite{bieri2013ElectromagneticAnalogGravitational}. We do so by exploring the Liénard--Wiechert solutions for the retarded potentials and fields of a moving charge. We can see how memory arises by considering the electromagnetic fields sourced by charges that are approximately inertial at early times, accelerate during intermediate times, and are again approximately inertial at late times.

    The text is structured as follows. To provide additional background, Sec. \ref{sec: background} discusses the various connections between the memory effect and modern research topics in infrared physics. Namely, we discuss in broad terms how memory is related to large gauge transformations and soft theorems. Throughout the text, we refer back to this section to gain new insights on the calculations we perform. Since Sec. \ref{sec: background} connects our main discussions to current research topics, some readers might prefer to skip it. The remainder of the paper can still be appreciated (and explored in a third-year undergraduate-level electrodynamics course) without it. In Sec. \ref{sec: LW-solution}, we briefly review the Liénard--Wiechert solutions for a single moving charge. In Sec. \ref{sec: field-lines}, these solutions are then used to illustrate how permanent changes in the electromagnetic field occur with the passage of radiation. This is done by drawing the electric field lines sourced by a charge that is inertial at early and late times, but accelerates during intermediate times. We also introduce the ``memory vector,'' given by the time integral of the electric field, and show how it relates to the change in the Coulombic field that is evident in the field lines. Section \ref{sec: large-gauge-transformations} focuses on the behavior of the vector potential, and in it, we discuss how the memory vector can be interpreted as a large gauge transformation. We also show how the vector potential in the inertial--accelerated--inertial setup behaves similarly to the sketch in Fig. \ref{fig: memory-pulse}. After the case of a single charge has been understood through the Liénard--Wiechert solution, we explain, in Sec. \ref{sec: radiation-from-scattering}, how that toy experiment can naturally arise in a scattering process. Once the main physical ideas have been laid out, we discuss in Sec. \ref{sec: experimental-prospects} how the memory effect could be probed in a laboratory. Lastly, we conclude in Sec. \ref{sec: conclusion} by briefly discussing the gravitational version of the memory effect, and the prospects for measuring it in modern gravitational wave observatories. The supplementary material contains a selection of solved exercises, the code used to produce Fig. \ref{fig: field-lines}, and code that can be used to reproduce Figs. \ref{fig: step-function} and \ref{fig: Ay-oscillating}.

    In the following, we mostly follow the conventions used by \textcite{griffiths2023IntroductionElectrodynamics}. In particular, three-dimensional vectors are denoted in boldface, and we work in SI units.

\section{Background}\label{sec: background}
    The memory effect is an exciting prediction on its own. It shows how even a hundred years of electrodynamics was not enough for us to learn all about its predictions. However, it also has many interesting connections with other topics in physics. Before we investigate memory quantitatively, it will be useful to have a bird's eye view of how it relates to various ideas.
    
    Let us first take a new look at the integral in Eq. \eqref{eq: kick-general-formula}. If we define the Fourier transform of the electric field as
    \begin{equation}
        \tilde{\vb{E}}(\omega,\vb{r}) = \int_{-\infty}^{+\infty} \vb{E}(t,\vb{r}) e^{i\omega t} \dd{t},
    \end{equation}
    then, for \(\omega = 0\),
    \begin{equation}
        \tilde{\vb{E}}(0,\vb{r}) = \int_{-\infty}^{+\infty} \vb{E}(t,\vb{r}) \dd{t}.
    \end{equation}
    This is the integral in Eq. \eqref{eq: kick-general-formula} for an infinite time interval. For an integral over a finite time interval, the frequency contributions are smeared into a packet, but the low frequencies dominate for large times. Since the low frequencies are the most important, we see that the memory effect probes the ``infrared structure'' of electrodynamics. Here, `infrared' means we are interested in what happens as we approach the limit of vanishing frequencies. It is also common to say that memory is a ``DC effect,'' in reference to direct current, because it is not associated with the ``large-frequency'' oscillations. 

    The frequency domain lets us look at memory in a different perspective, and it was through this perspective that a result equivalent to the memory effect was identified several years ago. It is a result in quantum field theory, and its modern formulation is due to \textcite{weinberg1965InfraredPhotonsGravitons}. 
    
    When computing observable quantities in quantum electrodynamics (for example, transition rates or cross sections), it is convenient to first compute a scattering amplitude---a complex quantity that stores information about the scattering process. \textcite{weinberg1965InfraredPhotonsGravitons} showed that the effects of very-low-energy (or ``soft'') photons can be easily accounted for by multiplying the expressions without the soft photons by an adequate ``soft factor.'' In symbols, one has
    \begin{equation}\label{eq: weinberg-soft-photon-theorem}
        A(\text{in} \to \text{out} + 1) = \frac{S_{\text{W}}}{\omega} A(\text{in} \to \text{out}) + o\qty(\frac{1}{\omega}).
    \end{equation}
    Above, \(A(\text{in} \to \text{out})\) is the scattering amplitude between some ``in'' state (such as a collection of incoming particles with known momenta) and some ``out'' state (such as a collection of outgoing particles with known momenta). In contrast, \(A(\text{in} \to \text{out} + 1)\) denotes the amplitude for the same process, but with the emission of an extra photon with frequency \(\omega\). \(S_{\text{W}}\) is the Weinberg soft factor, which corrects the calculation. Lastly, \(o\qty(1/\omega)\) denotes terms that are less important than the \(1/\omega\) term when \(\omega \to 0\) (i.e., when the photon is ``soft''). Together with the detailed expression for \(S_{\text{W}}\), Eq. \eqref{eq: weinberg-soft-photon-theorem} is known as the Weinberg soft photon theorem \cite{weinberg1965InfraredPhotonsGravitons}.

    The inverse Fourier transform of \(1/\omega\) is a step function in the time domain. As a consequence, the inverse Fourier transform of the scattering amplitude will have a step function component to it, akin to the function illustrated in Fig. \ref{fig: memory-pulse}. Together with the detailed expression of the Weinberg factor, this was shown to predict precisely how the soft photons yield the memory effect \cite{pasterski2017AsymptoticSymmetriesElectromagnetic}.
    
    Arguably, this makes Eq. \eqref{eq: kick-general-formula} even more interesting. Seen in the frequency domain, the memory effect is a consequence of very-low-frequency radiation, which means that the predicted shift in velocity is due to the effects of very-low-energy photons. It is surely unexpected that we can get physical observables from particles with near-zero energy---or, in the limiting case of infinite times, actual zero energy.

    The ease with which the Weinberg soft theorem allows us to account for very-low-energy photons suggests that there is a symmetry relating processes with and without soft photons. Symmetries come with conservation laws. Thus, there may be previously unknown conservation laws in electrodynamics. This turns out to be the case \cite{he2014NewSymmetriesMassless}. Weinberg's soft photon theorem is the expression, in quantum field theory, of conservation laws associated with ``large gauge transformations.'' These are gauge transformations that do not vanish at infinity. They are asymptotic symmetries of electrodynamics,  being ``asymptotic'' in the sense that they are important at very large distances. In the jargon, we say that the soft theorem is the Ward identity for the large gauge transformations.

    The cycle is complete when we relate large gauge transformations back to the memory effect. To do so, we consider what is meant by a ``vacuum'' in electrodynamics. Here, ``vacuum'' means a field configuration with no electromagnetic waves. Importantly, ``vacuum'' is not used here to refer to the absence of sources (i.e., charges and currents), but to the absence of electromagnetic waves. While large gauge transformations are a symmetry of quantum electrodynamics, they are not a symmetry of the vacua. In other words, there are multiple different vacua, which are related to each other by large gauge transformations. The memory effect is precisely a transition between these vacua. 

\begin{figure}[tb]
        \centering
        \includegraphics{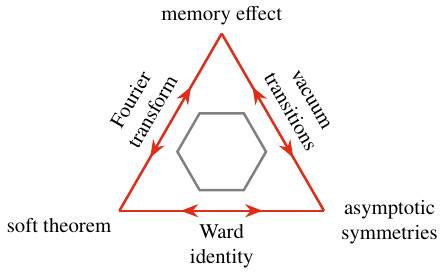}
        \caption{The infrared triangle. Memory effects are related to two other topics in infrared physics. The first are soft theorems, which are expressions in quantum field theory concerning how the addition of a very-low-energy particle affects a calculation. These expressions are essentially a reformulation of the memory effect in the frequency domain, and are thus related to memory through a Fourier transform. In turn, they are also the expressions, at the quantum level, of invariance under asymptotic symmetries. In the jargon, these types of expressions are known as Ward identities. Lastly, the memory effect can be understood as a transition between two vacua, both of them related by an asymptotic symmetry transformation. Figure adapted from Ref. \onlinecite{aguiaralves2025LecturesBondiMetzner}, which was published under a CC BY 4.0 license.}
        \label{fig: infrared-triangle}
    \end{figure}

    Together, these three topics (memory effects, soft theorems, and asymptotic symmetries) establish a triangle between different aspects of infrared physics. This is known as the infrared triangle, pictured in Fig. \ref{fig: infrared-triangle}.

    Many questions can arise at this point on various levels of technicality. 
    \begin{enumerate}
        \item Which physical scenario, if any, allows the integral in Eq. \eqref{eq: kick-general-formula} to be nonvanishing at order \(1/r\)?
        \item What is a vacuum in electrodynamics, and how can these different vacua be physically realized?
        \item Can the memory effect be measured in our universe at a finite distance, during a finite time, in a real laboratory? Or would it be necessary to carry a measurement during infinite time, or at an infinite distance, to see the effect? 
        \item Do very-low-energy particles (``soft photons'') really lead to physically observable effects, or are they automatically unphysical?
        \item If the memory effect is indeed physical, how is it related to gauge transformations?
    \end{enumerate}
    These are all very relevant questions for modern research on infrared physics. They can, in a sense, be summarized as ``Do these calculations imply observable predictions or should we dismiss them as unphysical mathematics?'' We will answer them throughout this article.

\section{The Liénard--Wiechert Solution}\label{sec: LW-solution}
    In describing electromagnetic phenomena, it is often useful to work with potentials. We will work in the Lorenz gauge, in which \cite[Eq. (10.12)]{griffiths2023IntroductionElectrodynamics}
    \begin{equation}
        \div\vb{A} = - \frac{1}{c^2}\pdv{V}{t},
    \end{equation}
    where \(V\) is the scalar potential, \(\vb{A}\) is the vector potential, and \(c\) is the speed of light. 
    
    In the Lorenz gauge, Maxwell's equations take the form of the wave equations
    \begin{subequations}\label{eq: wave-equation-potentials}
        \begin{align}
            - \frac{1}{c^2}\pdv[2]{V}{t} + \laplacian V &= - \frac{\rho}{\epsilon_0}, \\
            - \frac{1}{c^2}\pdv[2]{\vb{A}}{t} + \laplacian \vb{A} &= - \mu_0 \vb{J}, 
        \end{align}
    \end{subequations}
    where \(\rho\) is the charge density and \(\vb{J}\) is the current density.

    We will focus on the case in which the sources \(\rho\) and \(\vb{J}\) are given by a single pointlike charge \(Q\) following a known trajectory \(\vb{X}(t)\). We assume that this particle moves slower than light at all times, but it can achieve relativistic velocities. What are the electromagnetic fields sourced by this charge? The solution can be found in standard textbooks in the form of the Liénard--Wiechert potentials,
    \begin{subequations}\label{eq: LW-potentials}
    \begin{align}
        V(t,\vb{r}) &= \frac{\mu_0 c^2}{4 \pi} \frac{Q}{\alpha \norm{\vb{r} - \vb{X}(t_{\text{ret}})}}, \\
        \vb{A}(t,\vb{r}) &= \frac{\mu_0}{4 \pi} \frac{Q}{\alpha \norm{\vb{r} - \vb{X}(t_{\text{ret}})}}\dv{\vb{X}}{t}\qty(t_{\text{ret}}).
    \end{align}
    \end{subequations}
    Above, we denoted
    \begin{equation}\label{eq: def-alpha}
        \alpha = 1 - \frac{\vu{n}}{c} \vdot \dv{\vb{X}}{t}\qty(t_{\text{ret}}),
    \end{equation}
    where
    \begin{equation}\label{eq: def-n-vector}
        \vu{n} = \frac{\vb{r} - \vb{X}(t_{\text{ret}})}{\norm{\vb{r} - \vb{X}(t_{\text{ret}})}}.
    \end{equation}
    The retarded time \(t_{\text{ret}}\) is implicitly defined by 
    \begin{equation}\label{eq: def-tret}
        \norm{\vb{r} - \vb{X}(t_{\text{ret}})} = c (t - t_{\text{ret}}).
    \end{equation}

    These expressions contain much information. In particular, the retarded time \(t_{\text{ret}}\) occurs to take into account that moving a charge affects the electromagnetic field at a location \(\vb{r}\) only after \(t_{\text{ret}}-t\). In other words, the speed of light is finite. \(\vu{n}\) is the unit vector that points from the (retarded) position of the charge to the observation point, and \(\alpha\) encodes a Doppler-like correction to the expressions.

    Using the Liénard--Wiechert potentials, we can obtain the electric and magnetic fields by taking appropriate derivatives. The electric field is given by 
    \begin{multline}\label{eq: electric-field}
        \vb{E}(t,\vb{r}) = \frac{Q \mu_0 c^2}{4\pi}\qty(1 - \frac{1}{c^2}\norm{\dv{\vb{X}}{t}}^2) \frac{\qty(\vu{n} - \frac{1}{c}\dv{\vb{X}}{t})}{\alpha^3 \norm{\vb{r} - \vb{X}(t_{\text{ret}})}^2} \\ + \frac{Q \mu_0}{4\pi}\frac{\vu{n} \cp \qty[\qty(\vu{n} - \frac{1}{c}\dv{\vb{X}}{t}) \cp \dv[2]{\vb{X}}{t}]}{\alpha^3 \norm{\vb{r} - \vb{X}(t_{\text{ret}})}},
    \end{multline}
    where the derivatives \(\dv*{\vb{X}}{t}\) and \(\dv*[2]{\vb{X}}{t}\) are evaluated at \(t_{\text{ret}}\). The magnetic field is
    \begin{equation}\label{eq: B-n-E-LW-solution}
        c \vb{B}(t,\vb{r}) = \vu{n} \cp \vb{E}(t,\vb{r}).
    \end{equation}
    This solution can be used to justify the small velocity approximation we did in Eq. \eqref{eq: acceleration-test-charge-for-kick}; see Exercise S.1 in the supplementary material.

    Equation \eqref{eq: electric-field} is the general expression for the electric field sourced by a moving charge. It neatly separates into two components with very different properties. They are the following.
    \begin{enumerate}
        \item The ``Coulombic field'' is the first line of Eq. \eqref{eq: electric-field}. It is a generalization of the Coulomb law found in electrostatics and it falls off as \(1/r^2\). This term depends on the velocity of the source and includes Doppler-like corrections, but it does not depend on the acceleration. 
        \item The ``radiative field'' is the second line of Eq. \eqref{eq: electric-field}. This is a feature that cannot be found in electrostatics because it is proportional to the source's acceleration. Furthermore, it falls off merely as \(1/r\), which means that it dominates the Coulombic field sufficiently far away from the source. 
    \end{enumerate}
    In a sense, the Coulombic field is attached to the source, while the radiative field is not. For example, the radiative field carries energy away from the source and exhibits how the electromagnetic field is a physical entity of its own, rather than being ``created'' by the source (see Chap. 1 in Ref. \onlinecite{wald2022AdvancedClassicalElectromagnetism}).

    While Eq. \eqref{eq: electric-field} allows us to clearly distinguish a Coulombic component and a radiative component, we cannot do the same when looking at the Liénard--Wiechert potentials in Eq. \eqref{eq: LW-potentials}. Both the Coulombic and the radiative part merge into the same \(1/r\)-decay. This coincidence leads to very interesting consequences. For example, suppose we have a charged particle that initially moves inertially, then undergoes acceleration, and eventually returns to inertial motion. We then have three stages.
    \begin{enumerate}
        \item At early times, the particle moves inertially, and thus the electromagnetic field is given solely by the Coulombic contribution. The radiative field vanishes. 
        \item At intermediate times, the particle undergoes an acceleration, and thus the radiative field is nonvanishing. There is still the always-present Coulombic contribution.
        \item At late times, the particle moves inertially again, and the radiative field vanishes. We have only a Coulombic contribution. However, the velocity of the source may differ from what it was at early times. If this is the case, the Coulombic field will also differ from its value at early times. 
    \end{enumerate}

    This structure gives a first hint of how memory can arise in the Liénard--Wiechert fields. As discussed in Sec. \ref{sec: background}, the memory effect can be understood as a change of vacuum, where ``vacuum'' means a field configuration without electromagnetic waves. In the Liénard--Wiechert solution, this means the radiative field should vanish. There can be multiple different configurations that are purely Coulombic, each sourced by a particle moving by a different constant velocity. Therefore, we can have different vacua. Hence, if we want to ``generate memory,'' we can start by considering a source that follows an inertial--accelerated--inertial trajectory, but has a finite change in its velocity between early and late times. This will lead to different Coulombic fields between early and late times and could lead to memory. 

\begin{figure}[tb]
        \centering
        \includegraphics{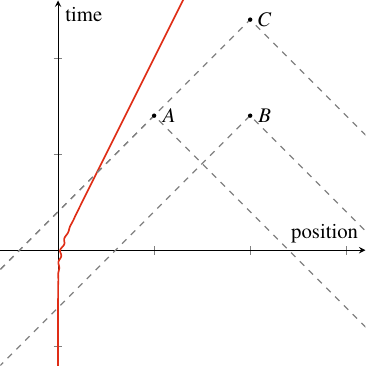}
        \caption{Minkowski diagram illustrating the effects of moving away from the source at constant laboratory time \(t\) or at constant retarded time \(t_{\text{ret}}\). The dark curve illustrates the motion of the source charge. The dashed lines indicate light cones---positive slope dashed lines have constant retarded time \(t_{\text{ret}}\). Let us use measurements at the event \(A\) (i.e., at the position and time of \(A\)) as a reference. If we move farther away from the source, relative to \(A\), at constant \(t\), we reach \(B\). Notice the light cone of \(B\) intersects the trajectory of the particle before the accelerated stage, so before any radiation was emitted. In contrast, if we move away from the source, relative to \(A\), along a direction of constant \(t_{\text{ret}}\), we reach \(C\). The light cones of \(A\) and \(C\) intersect the particle's trajectory at the same point, and hence both have already seen the emission of radiation. Nevertheless, \(C\) is farther away, and the Coulombic contribution is less relevant.}
        \label{fig: trajectory-finite-velocity}
    \end{figure}

    Before proceeding, we should clarify an important subtlety in what it means for the Coulombic and radiative fields to fall off as \(1/r^2\) or \(1/r\). When we say ``the radiative field falls off as \(1/r\),'' we mean that 
    \begin{equation}
        \lim_{r \to \infty} r \vb{E}_{\text{rad}}
    \end{equation}
    is finite. Above, \(\vb{E}_{\text{rad}}\) stands for the radiative part of the Liénard--Wiechert solution. Notice, though, that taking this limit requires us to say what other variables are being held constant. Notably, we can choose to keep \(t\) constant, or we can choose to keep \(t_{\text{ret}}\) constant. These two choices lead to different results! If we take the limit at constant \(t_{\text{ret}}\), the dominant term is the radiative term, and we are reaching the radiation zone. However, if we take the limit at constant \(t\), then large values of \(r\) correspond to very negative values of \(t_{\text{ret}}\). This means that the field at large \(r\) and constant \(t\) was sourced at early times. In the inertial--accelerated--inertial setup, this means the source charge had not accelerated yet. Hence, the radiative field vanishes, and the sole contribution is Coulombic. This is illustrated with a spacetime diagram in Fig. \ref{fig: trajectory-finite-velocity}.

    For this reason, whenever we talk about the large distance behavior in this article, we mean ``large \(r\) at constant retarded time.'' This means that, sometimes, it will not be convenient to use laboratory time \(t\) in the Maxwell equations. Since \(t_\text{ret}\) depends a lot on the particle trajectory through Eq. \eqref{eq: def-tret}, we can instead use the ``retarded time to the origin,''
    \begin{equation}\label{eq: def-u}
        u = t - \frac{r}{c}.
    \end{equation}
    Equation \eqref{eq: def-tret} implies that 
    \begin{equation}\label{eq: u-t-ret-rel}
        u = t_{\text{ret}} + \frac{\norm{\vb{r} - \vb{X}(t_{\text{ret}})} - r}{c}.
    \end{equation}
    Note that, at large \(r\),
    \begin{equation}
        u=t_\text{ret}-\frac{\vu{r}\vdot\vb{X}(t_\text{ret})}{c}+\order{\frac{1}{r}}
    \end{equation}
    remains finite when \(t_\text{ret}\) is fixed. In fact, notice that \(u\) and \(t_{\text{ret}}\) differ by a finite amount in that limit. While their difference depends on the direction we are looking at, it does not depend on \(r\). For this reason, taking large \(r\) limits with \(u\) held constant is equivalent to taking large \(r\) limits with \(t_{\text{ret}}\) held constant.

\section{Field Lines}\label{sec: field-lines}
    Our natural next step is then to investigate the change in the Coulombic field between early and late times. We can see this change without using the Liénard--Wiechert solutions in their full glory by considering the behavior of the electric field lines. Although the fields at intermediate times can be extremely complicated, we know that at early and late times they are Coulombic. This implies, as we will see, that they are much easier to deal with.
    
    Electric field lines must obey, among others, the following rules: 
    \begin{enumerate}
        \item Field lines can never cross each other (fields must be single valued), and \label{item: field-lines-never-cross}
        \item Their flux through any closed surface measures the enclosed electric charge (Gauss's law). \label{item: flux-through-closed-surface}
    \end{enumerate}
    Using these rules and the assumption that the sources move inertially at early and late times, we can already sketch the appearance of the electric field lines.

    At early and late times, the sources move inertially. If we substitute \(\vb{X}(t) = \vb{X}(0) + \vb{v}t\) in Eq. \eqref{eq: electric-field}, we find
    \begin{equation}\label{eq: electric-field-inertial}
        \vb{E}(t,\vb{r}) = \frac{Q \mu_0 c^2}{4\pi}\qty(1 - \frac{v^2}{c^2}) \frac{\qty(\vu{n} - \vb{v}/c)}{(1-\vu{n}\vdot\vb{v}/c)^3 \norm{\vb{r} - \vb{X}(t_{\text{ret}})}^2},
    \end{equation}
    with \(v = \norm{\vb{v}}\) for simplicity. We also expanded the definition of \(\alpha\) according to Eq. \eqref{eq: def-alpha}.
    
\begin{figure}[tb]
        \centering
        \includegraphics{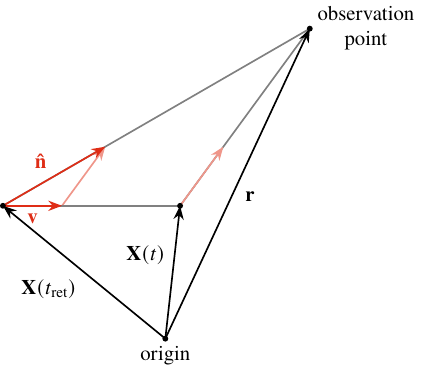}
        \caption{For an inertially moving charge with velocity \(\vb{v}\), the combination \(\vu{n}-\vb{v}/c\) is proportional to \(\vb{r}-\vb{X}(t)\). This means the electric field points to the charge's current position, not to its retarded position. Notice that the distance between the observation point and the charge's retarded position is \(c(t-t_{\text{ret}})\) due to the definition of retarded time. Meanwhile, the distance between the charge's retarded and current positions is \(v(t-t_{\text{ret}})\) due to the inertial movement. The diagram sets \(c = 1\) for ease of visualization.}
        \label{fig: retarded-position}
    \end{figure}

    Equation \eqref{eq: electric-field-inertial} surprisingly states that the electric field lines point to the position of the source at time \(t\), not \(t_\text{ret}\). This is because
    \begin{subequations}
        \begin{align}
            \vu{n} - \frac{\vb{v}}{c} &= \frac{\vb{r} - \vb{X}(t_{\text{ret}})}{\norm{\vb{r} - \vb{X}(t_{\text{ret}})}} - \frac{\vb{X}(t) - \vb{X}(t_{\text{ret}})}{c (t - t_{\text{ret}})}, \\
            &= \frac{\vb{r} - \vb{X}(t)}{c (t - t_{\text{ret}})},
        \end{align}
    \end{subequations}
    where we used Eqs. \eqref{eq: def-n-vector} and \eqref{eq: def-tret}. This result is illustrated in Fig. \ref{fig: retarded-position}.

\begin{figure}[tb]
        \centering
        \includegraphics{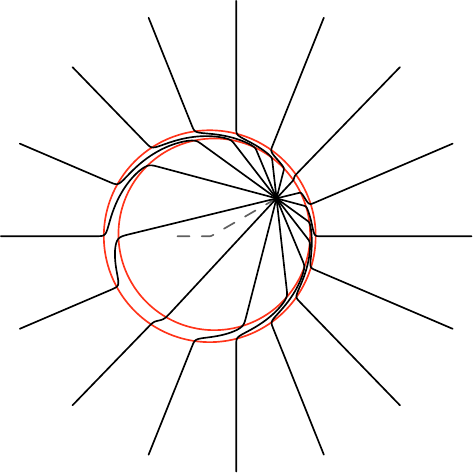}
        \caption{Electric field lines for the Liénard--Wiechert solution. In this illustration, a particle initially moves horizontally at \(v=0.25c\), then undergoes acceleration for a short period of time, and finally moves on inertially at \(v=0.75c\) and at an angle of \(\ang{30}\) with the horizontal axis. The trajectory is illustrated by a dashed line in the background. The dashed line begins at the position the charge was at when the outermost points depicted in the field lines were ``emitted.'' In the diagram, the radiation due to the charge's acceleration is (mostly) enclosed by the two highlighted circles, outside of which the charge is nearly inertial. Notice how the field lines at early times (outside the circles) are much closer to being isotropic than those at late times (inside the circles). This figure was drawn using \textsc{Asymptote} \cite{bowman2008AsymptoteVectorGraphics,hammerlindl2004AsymptoteDescriptiveVector}, and the code is available as part of the supplementary material.}
        \label{fig: field-lines}
    \end{figure}

    If the electric field due to an inertial source points to the present position of the source, we can already sketch the field lines for the inertial--accelerated--inertial setup described above. At late times, the field points directly to the present position of the source. At early times, it points to the would-be position of the source if it kept moving inertially. Rule \ref{item: flux-through-closed-surface} forces us to draw the same number of field lines close to the charge and at far distances from it. Rule \ref{item: field-lines-never-cross} demands that we connect the inner adjacent lines to the outer adjacent lines. Hence, we learn that the electric field lines of this setup should resemble those in Fig. \ref{fig: field-lines}.
    
    An inspection of Fig. \ref{fig: field-lines} reveals a change in the Coulombic field at early and late times. The field due to the early-time motion is seen outside the highlighted circles. The source was moving at relatively low speeds at that time, and therefore, the field lines look almost isotropic. At late times, we get the field lines inside the innermost circle. The charge is then moving very fast, and, as a consequence, the field lines distinctively distinguish between the direction of motion and the direction orthogonal to it. While we have a Coulombic field at both early and late times, the two Coulombic fields differ. This difference originates at intermediate times, when the acceleration of the charge leads to radiation. The coincidence of radiative and Coulombic orders in the potentials implies that this radiation is capable of leaving a lasting change in the Coulombic field.

    Note that, since the Coulombic field is the part of the Liénard--Wiechert solution that is void of electromagnetic waves, the Coulombic configurations shown in Fig. \ref{fig: field-lines} are precisely what we called `vacua' in Sec. \ref{sec: background}. Figure \ref{fig: field-lines} is an example of how a vacuum transition can look like in electrodynamics.

    \subsection{Memory and changes in the Coulombic field}\label{subsec: memory-changes-coulombic}
        At this point, it is natural to inquire whether the change in the Coulombic field depicted in Fig. \ref{fig: field-lines} is indeed related to the memory effect. The answer is positive. To prove this, we need to show that a change in the Coulombic field is somehow related to the integral in Eq. \eqref{eq: kick-general-formula}. For concreteness, let us define the ``memory vector'' \(\vb*{\Delta}\) as
        \begin{equation}\label{eq: memory-vector}
            \vb*{\Delta}(\vb{r}) = \int_{-\infty}^{+\infty} \vb{E}(t,\vb{r}) \dd{t}.
        \end{equation}
        Equation \eqref{eq: kick-general-formula} means the memory effect is completely encoded in \(\vb*{\Delta}\). A quantity similar to \(\vb*{\Delta}\) was first introduced by \textcite{satishchandran2019AsymptoticBehaviorMassless} when studying the memory effect in general relativity.

        The memory vector is subtle. While the Coulombic field contributes terms that can become infinite, our interest lies in the radiative sector. We therefore concentrate our attention on the \(1/r\) behavior, which characterizes the radiation field and its contribution to the memory vector. As explained at the end of Sec. \ref{sec: LW-solution}, the ``\(1/r\) behavior'' is meant at constant \(t_{\text{ret}}\), not at constant \(t\). Equivalently, we can work with the ``retarded time to the origin'' \(u = t - r/c\) instead of \(t_{\text{ret}}\). Since the radiative part of the Liénard--Wiechert solution falls off as \(1/r\), but the Coulombic part falls off as \(1/r^2\), we see that
        \begin{equation}\label{eq: electric-field-large-r}
            \vb{E} = \vb{E}_{\text{rad}} + \order{\frac{1}{r^2}},
        \end{equation}
        where \(\vb{E}_{\text{rad}}\) denotes the radiative part of the Liénard--Wiechert solution. Therefore,
        \begin{equation}\label{eq: memory-vector-LW}
            \vb*{\Delta}(\vb{r}) = \int_{-\infty}^{+\infty} \vb{E}_{\text{rad}} \dd{u} + \order{\frac{1}{r^2}},
        \end{equation}
        where we also made a change of variables in the integral to ensure the large-\(r\) limits are taken correctly.
    
        Let us now consider the Gauss law. We have 
        \begin{equation}
            \qty(\div\vb{E})_t = \frac{\rho}{\epsilon_0},
        \end{equation}
        where the divergence is computed at constant \(t\) (indicated by the parentheses). Since we are interested in the large-\(r\) behavior at constant \(u = t-r/c\), let us use the chain rule to rewrite this expression. We find (Exercise S.2 in the supplementary material)
        \begin{equation}
            \qty(\div\vb{E})_u - \frac{\vu{r}}{c} \vdot \qty(\pdv{\vb{E}}{u})_{\vb{r}} = \frac{\rho}{\epsilon_0}.
        \end{equation}
        with the divergence now computed at constant \(u\). Upon integrating over \(u\), we obtain
        \begin{equation}
            \vu{r}\vdot\Delta \vb{E} = c \int \div{\vb{E}} \dd{u} - \frac{c}{\epsilon_0} \int \rho \dd{u}.
        \end{equation}
        \(\Delta \vb{E}\) is the change in the electric field before and after the pulse. Therefore, it is Coulombic, and thus falls off as \(\order{1/r^2}\). Only the radiative field will contribute to \(\div\vb{E}\) as \(\order{1/r^2}\). Hence, we can write
        \begin{equation}
            \vu{r}\vdot\Delta \vb{E} = c \int \div{\vb{E}_{\text{rad}}} \dd{u} - \frac{c}{\epsilon_0} \int \rho \dd{u} + \order{\frac{1}{r^3}}.
        \end{equation}

        The sources we are considering are pointlike charges. Hence, \(\rho\) cannot extend to infinity, and it vanishes at large \(r\). We may then write 
        \begin{equation}
            \vu{r}\vdot\Delta \vb{E} = c \int \div{\vb{E}_{\text{rad}}} \dd{u} + \order{\frac{1}{r^3}}.
        \end{equation}
        For the Liénard--Wiechert solution, we can use Eq. \eqref{eq: memory-vector-LW} to conclude that
        \begin{equation}\label{eq: radial-coulombic-from-memory-vector}
            \vu{r}\vdot\Delta \vb{E} = c \div{\vb*{\Delta}} + \order{\frac{1}{r^3}}.
        \end{equation}
        This relates the leading order change in the Coulombic field to the memory vector. In other words, the change in the Coulombic field depicted in Fig. \ref{fig: field-lines} is indeed a signal of the memory effect. At leading order, it shows us that the memory vector does not vanish. This means that a test charge will undergo a finite velocity kick upon the passage of the wave pulse.

        It is worth noticing that this change in the field lines before and after the passage of radiation has been well-known for over a hundred years. The procedure outlined above to sketch the field lines near a fast burst of radiation can be traced back to \textcite{[{}][{, Chap. 3.}]{thomson1904ElectricityMatter}}, and the Liénard--Wiechert solutions are just as old \cite{lienard1898ChampElectriqueMagnetique,*lienard1898ChampElectriqueMagnetique2,*lienard1898ChampElectriqueMagnetique3,wiechert1901ElektrodynamischeElementargesetze}. Nevertheless, noticing this also leads to a velocity kick on the test particles is much more recent \cite{bieri2013ElectromagneticAnalogGravitational}. 

\section{Vector Potential}\label{sec: large-gauge-transformations}
    In Sec. \ref{subsec: memory-changes-coulombic}, we have established that Fig. \ref{fig: field-lines} is a depiction of the memory effect. More precisely, in the setup illustrated by Fig. \ref{fig: field-lines}, the memory vector is nonvanishing. However, our discussion required doing expansions at large \(r\), and we may wonder whether it would actually be possible to distinguish the memory signal in a laboratory. In practice, we cannot measure only the \(1/r\) term without measuring the \(1/r^2\) term as well, so is the memory kick observable? To assess this, it will be convenient to choose a specific model for the trajectory of the source particle. We will also take this opportunity to explore the behavior of the potentials. 

    Let us consider some step-like function \(H(t)\). We ask it to be smooth, approximately \(0\) for negative values of \(t\), and approximately \(1\) for positive values of \(t\). Two possible examples are shown in Fig. \ref{fig: step-function}, and further detailed in Exercise S.3 in the supplementary material.

    \begin{figure}[tb]
        \centering
        \includegraphics{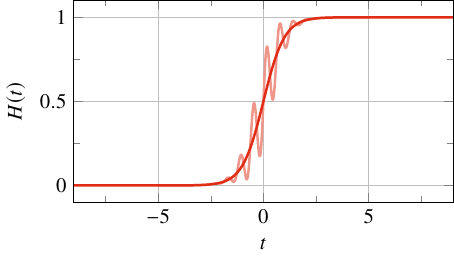}
        \caption{Examples of functions smoothly rising from zero to unit. The function may oscillate while rising. These examples are two particular cases of the functions \(H_{\omega}(t)\) defined on Exercise S.3 in the supplementary material. Namely, \(\omega = 0\) and \(\omega = 10\).}
        \label{fig: step-function}
    \end{figure}

    We can use such a function \(H(t)\) to write an interesting velocity profile for a source charge. Define
    \begin{equation}\label{eq: velocity-profile-step}
        \dot{\vb{X}}(t) = \vb{v}_{\text{in}} + H\qty(\frac{t}{T}) \qty(\vb{v}_{\text{out}} - \vb{v}_{\text{in}}),
    \end{equation}
    where \(T\) is some time scale controlling how long the intermediate times last. If \(H(t)\) is carefully chosen, the velocity of the source never exceeds the speed of light (Exercise S.4 in the supplementary material). Note also that for \(t \ll -T\), \(\dot{\vb{X}}(t) \approx \vb{v}_{\text{in}}\) (early times), while for \(t \gg T\) we have \(\dot{\vb{X}}(t) \approx \vb{v}_{\text{out}}\) (late times). The acceleration happens for time \(t\) such that \(-T \lesssim t \lesssim T\). A velocity profile like Eq. \eqref{eq: velocity-profile-step} gives a simple way to model a source that initially moves at constant velocity, undergoes some complicated acceleration phase, and eventually moves on at a different constant velocity.

\begin{figure}[tb]
        \centering
        \includegraphics{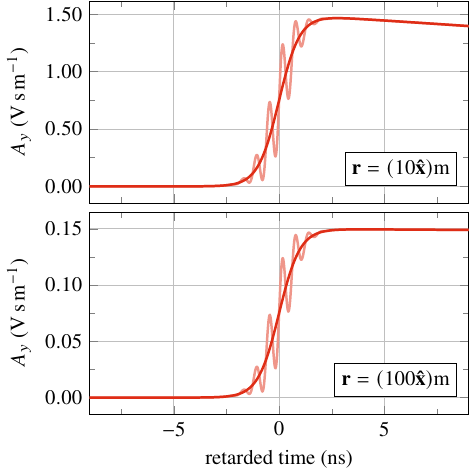}
        \caption{Nonvanishing component of the vector potential for the Liénard--Wiechert solution sourced by a charge with velocity given by Eq. \eqref{eq: velocity-profile-step}. We show the computations for the same choices of function \(H(t)\) illustrated on Fig. \ref{fig: step-function}. This solution has \(\vb{v}_{\text{in}} = \vb{0}\), \(\vb{v}_{\text{out}} = (0.5 \vu{y}) c\), \(T = \SI{1}{\nano\second}\) and considers a source with charge \(\SI{1}{\coulomb}\). Since the potential depends linearly on the source charge, the potential sourced by smaller charges can be easily obtained by rescaling the vertical axis. The graph depicts the time-dependence of the \(\vu{y}\) component of \(\vb{A}\) at the observation point \(\vb{r} = (10 \vu{x})\si{\meter}\) (top panel) and at the observation point \(\vb{r} = (100 \vu{x})\si{\meter}\) (bottom panel). Note that the value of the vector potential has a lasting shift after the passage of the pulse, akin to the sketch in Fig.~\ref{fig: memory-pulse}. Note also that the slope seen at late times is much smaller in the bottom panel than at the top panel. As explained in the text, this slope is a finite-distance effect.}
        \label{fig: Ay-oscillating}
    \end{figure}

    Using Eq. \eqref{eq: velocity-profile-step} with a specific choice of \(H(t)\), we can obtain \(\vb{X}(t)\) through direct integration after providing an initial condition \(\vb{X}(0)\). With \(\vb{X}(t)\) in hand, the Liénard--Wiechert formulae determine the potentials everywhere in spacetime. Figure \ref{fig: Ay-oscillating} depicts them at two particular observation points. The exhibited component of the vector potential starts at zero at early times, oscillates during intermediate times, and eventually reaches a new value at late times. This new value is not completely constant. However, the time variation is much slower compared to intermediate times. The top panel of Fig. \ref{fig: Ay-oscillating} almost resembles our illustration of a pulse with memory in Fig. \ref{fig: memory-pulse}---if only we did not have this curious slope at late times! The bottom panel has no slope, and resembles Fig. \ref{fig: memory-pulse} much more closely. 

    Figure \ref{fig: Ay-oscillating} can be understood from the Liénard--Wiechert expression in Eq. \eqref{eq: LW-potentials}. At early times, the particle is stationary, and the vector potential vanishes, in accordance with our expectation from electrostatics. Later, as the velocity changes following Eq.~\eqref{eq: velocity-profile-step}, the vector potential follows along, because it is proportional to the velocity. At late times, the velocity reaches a new fixed value. Hence, all the time dependence left in the vector potential is due to the factor \(\norm{\vb{r} - \vb{X}(t_{\text{ret}})}\) in the denominator. This factor forces the potential to diminish as the source moves away from the observation point. Since \(\vb{X}(t_{\text{ret}})\) varies linearly with time at this stage, the vector potential varies at a much slower rate than it did at intermediate times. This is the origin of the slope at late times depicted in the top panel of Fig. \ref{fig: Ay-oscillating}, which is absent in Fig. \ref{fig: step-function}. Could this behavior spoil a laboratory measurement of the memory effect?

    Let us take a closer look at the time variation of the vector potential at late times by considering the Liénard--Wiechert potential \eqref{eq: LW-potentials} for an inertial particle. It gives, upon expanding the definitions of \(\vu{n}\) and \(\alpha\),
    \begin{equation}
        \vb{A}(t,\vb{r}) = \frac{\mu_0}{4 \pi} \frac{Q c \vb{v}}{c \norm{\vb{r} - \vb{v} t_{\text{ret}}} - (\vb{r} - \vb{v} t_{\text{ret}})\vdot \vb{v}}.
    \end{equation}
    All time dependence in this expression comes through the retarded time \(t_{\text{ret}}\), which is defined by Eq. \eqref{eq: def-tret}. For the model \eqref{eq: velocity-profile-step}, the wave front associated with the acceleration of the source arrives at an observation point at approximately \(t_{\text{ret}} \in (-T, T)\). This is the case because the source was undergoing acceleration near the origin during time \(t \in (-T,T)\). Hence, the ``interesting'' dynamics of the vector potential happens at finite values of \(t_{\text{ret}}\). Meanwhile, at very large distances, we approximate
    \begin{equation}\label{eq: large-r-retarded-time}
        r \gg c \abs{t_{\text{ret}}}.
    \end{equation}
    This approximation is justified precisely because we are interested in finite values of \(t_{\text{ret}}\). The vector potential then becomes
    \begin{equation}\label{eq: vector-potential-inertial}
        \vb{A}(t,\vb{r}) = \frac{\mu_0}{4 \pi r} \frac{Q \vb{v}}{(1 - \vu{r}\vdot\vb{v}/c)} + \order{\frac{1}{r^2}}.
    \end{equation}
    Hence, to the leading order in \(r\), the vector potential is constant in time. This means that the slope we see at late times in the top panel of Fig. \ref{fig: Ay-oscillating} is due to finite-distance effects. If we picked an observation point farther away from the source, the slope would be even smaller. This is shown in the bottom panel of Fig. \ref{fig: Ay-oscillating}. As expected, the slope is now too small to be noticed. 

    The approximation in Eq. \eqref{eq: large-r-retarded-time} may appear artificial, but it becomes more natural if we write it in a different way. Notice that, since \(\norm{\vb{v}} < c\), we may also write
    \begin{equation}
        \norm{\vb{r}} \gg \norm{\vb{v} t_{\text{ret}}}.
    \end{equation}
    This same equation can be written as 
    \begin{equation}
        \norm{\vb{r}} \gg \norm{\vb{X}(t_{\text{ret}})}.
    \end{equation}
    Hence, Eq. \eqref{eq: large-r-retarded-time} implies that we are interested in the radiative behavior far away from the source. Nevertheless, for a source that is able to move arbitrarily far away, this assumption will inevitably break down if we sit at a finite \(\vb{r}\) and wait long enough.

    The upshot is that we can hope to measure memory. In Sec. \ref{sec: field-lines}, we discussed a ``large \(r\) expansion'' without being very precise about how large \(r\) needed to be. Now, however, we have Eq. \eqref{eq: large-r-retarded-time} giving a much more precise comparison. In the top panel of Fig. \ref{fig: Ay-oscillating}, we are at \(r = \SI{10}{\meter}\) from the source, but we have \(c t_{\text{ret}} \approx \SI{1.5}{\meter}\) (using \(t_{\text{ret}} \approx \SI{5}{\nano\second}\)). This is good, but still gives us the curious slope. If we move farther away, we can easily claim that \(r = \SI{100}{\meter} \gg \SI{1.5}{\meter} \approx c t_{\text{ret}}\), and we no longer see any slope. Even though the memory effect relates to particles with very-low-energies, at very large distances, and in other ``extreme'' scenarios, Fig. \ref{fig: Ay-oscillating} shows that we should still be able to see it at a finite distance from the source. The price for that is set by Eq. \ref{eq: large-r-retarded-time}, which tells us that we would need to perform our measurements fast. We shall return to this point in Sec. \ref{sec: experimental-prospects}.

    \subsection{Memory and changes in the vector potential}
        We have left a loose end in the previous paragraphs. While Figs. \ref{fig: memory-pulse} and \ref{fig: Ay-oscillating} resemble each other, we have not yet proved that Fig. \ref{fig: Ay-oscillating} is an actual illustration of the memory effect. More precisely, we have not shown yet whether the shift in the vector potential between early and late times is mathematically related to the memory vector defined on Eq. \eqref{eq: memory-vector}. Let us then pursue and understand this connection.
        
        Directly from the Liénard-Wiechert potentials (and also in more generality, see Sec. 20.5.4 in Ref. \onlinecite{zangwill2013ModernElectrodynamics}), we have
        \begin{equation}\label{eq: electric-field-large-r-A}
            \vb{E}(t,\vb{r}) = \vu{r} \cp \qty[\vu{r} \cp \pdv{\vb{A}}{t}] + \order{\frac{1}{r^2}}.
        \end{equation} 
        Plugging into Eq. \eqref{eq: memory-vector},
        \begin{equation}
            \vb*{\Delta} = \vu{r} \cp \qty[\vu{r} \cp \Delta \vb{A}] + \order{\frac{1}{r^2}}.
        \end{equation}
        To the leading order, we can rewrite the triple vector product as 
        \begin{subequations}\label{eq: memory-vector-potential}
            \begin{align}
                \vb*{\Delta} &= (\vu{r} \vdot \Delta \vb{A}) \vu{r} - \Delta \vb{A} + \order{\frac{1}{r^2}}, \\
                &= - [\Delta \vb{A}]^{\text{T}} + \order{\frac{1}{r^2}}.
            \end{align}
        \end{subequations}
        The ``T'' superscript stands for ``transverse,'' because the combination \(\Delta \vb{A} - (\vu{r} \vdot \Delta \vb{A}) \vu{r}\) equals ``\(\Delta \vb{A}\) with its radial component removed.'' This reinforces that the memory vector is transverse at leading order---which was already known because its leading behavior depends only on the radiative electric field---and that it corresponds to the leading change in the vector potential. 
        
        Note that we are working in Lorenz gauge. Equation \eqref{eq: electric-field-large-r-A} holds in Lorenz gauge, and can be shown to hold in several other related gauges as well (Exercise S.5 in the supplementary material). It follows that we could repeat the derivation in these other gauges, and conclude that Eq. \eqref{eq: memory-vector-potential} is gauge-invariant! This is possible because Eq. \eqref{eq: memory-vector-potential} is concerned only with a \emph{difference} of the \emph{transverse components} of the \(1/r\) piece of the vector potential. Trying to gauge \(\Delta \vb{A}\) away would require us to change the value of \(\pdv*{\vb{A}}{t}\) at intermediate times. But the \(1/r\) piece of the transverse components of \(\pdv*{\vb{A}}{t}\) are precisely what contribute to the electric field in Eq. \eqref{eq: electric-field-large-r-A}. In fact, while we chose to work in Lorenz gauge, a popular choice in the literature is instead the radial gauge (\(\vb{A} \vdot \vu{r} = 0\))---see, e.g., Ref. \onlinecite{pasterski2017AsymptoticSymmetriesElectromagnetic}. Physically, Eq. \eqref{eq: memory-vector-potential} explicitly relates the shift in the vector potential to the memory vector. Therefore, it establishes that the shift in the vector potential has the same physical origin as the velocity kick that would be imparted on a test charge.

        Let us compute the curl of the memory vector. We find
        \begin{subequations}
            \begin{align}
                \curl\vb*{\Delta} &= \int_{-\infty}^{+\infty} \curl\vb{E} \dd{t}, \\
                &= - \int_{-\infty}^{+\infty} \pdv{\vb{B}}{t} \dd{t}, \label{eq: curl-memory-faraday-step} \\
                &= \left.\vb{B}\right|_{-\infty} - \left.\vb{B}\right|_{+\infty},
            \end{align}
        \end{subequations}
        where we used Faraday's law in Eq. \eqref{eq: curl-memory-faraday-step}. For the Liénard--Wiechert solution of inertially moving charges, the magnetic field will decay at least as \(1/r^3\). Hence, 
        \begin{equation}\label{eq: curl-memory-vector}
            \curl\vb*{\Delta} = \vb{0} + \order{\frac{1}{r^3}}.
        \end{equation}
        Therefore, Helmholtz's theorem tells us there is a scalar function \(\lambda\) decaying as \(\order{r^0}\) such that
        \begin{equation}
            \vb*{\Delta} = -\grad\lambda + \order{\frac{1}{r^2}}.
        \end{equation}
        Since \(\vb*{\Delta}\) is time-independent, so is \(\lambda\). Therefore, Eq. \eqref{eq: memory-vector-potential} allows us to conclude that the difference in the potentials at early and late times obeys
        \begin{subequations}\label{eq: change-potentials-large-gauge-transformation}
            \begin{align}
                \Delta V &= - \pdv{\lambda}{t} + \order{\frac{1}{r^2}}, \\
                \Delta \vb{A} &= \grad \lambda + \order{\frac{1}{r^2}}, 
            \end{align}
        \end{subequations}
        where we exploited the fact that \(\pdv*{\lambda}{t}\) and \(\pdv*{\lambda}{r}\) vanish. This means that the leading change in the potentials between early and late times is mathematically identical to a gauge transformation! By carefully picking the subleading terms, we can ensure that this gauge transformation does not break the Lorenz gauge. Because \(\lambda\) is of the order \(\order{r^0}\), and thus does not vanish at infinity, it is said to be a large gauge transformation \cite{he2014NewSymmetriesMassless}.

        What is the meaning of this gauge transformation? Suppose that, at early times, we choose the Lorenz gauge such that the vacuum (absence of electromagnetic waves) is defined by \(\vb{A} = \vb{0}\). Then Eq. \eqref{eq: change-potentials-large-gauge-transformation} tells us that, at late times, the new vacuum will be given by \(\vb{A} = \grad\lambda\). In other words, we have established that the large gauge transformation is the connection between the two vacua depicted in Fig. \ref{fig: field-lines}.

\section{Radiation from Scattering}\label{sec: radiation-from-scattering}
    In Secs. \ref{sec: field-lines} and \ref{sec: large-gauge-transformations}, we extensively studied how the movement of a single charge can leave a permanent imprint in the electromagnetic field. Then, we had assumed that the particle followed some complicated trajectory that was inertial at early times, accelerated at intermediate times, and inertial again at late times. In order to obtain a permanent imprint---or memory---the velocity of this source charge at early and late times had to be different. We now set out to find physical scenarios in which this can happen realistically. 

    Using a single charge, it is difficult to produce acceleration. Acceleration happens through the action of a force, which means that the charge interacts with something. If we need the charge to interact with something, we may simply add more charges and let them interact! Since the Maxwell equations are linear in the electromagnetic fields, we can just solve for the Liénard--Wiechert fields of each particle and add the solutions together to obtain the complete field. Since the charges respond to the fields sourced by each other, they now undergo acceleration in a natural way. 

\begin{figure}[tb]
        \centering
        \includegraphics{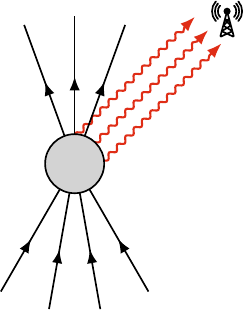}
        \caption{A number of charged particles approach each other. At early times, they move approximately inertially. They interact nonnegligibly for a finite time, during which they may emit electromagnetic radiation. After that time, a number of particles are sent off in approximately inertial trajectories. The particles are allowed to merge or split during the interaction, and thus the total number of particles does not need to be conserved. The electromagnetic radiation resulting from the process may then be detected far away---for example, in an antenna.}
        \label{fig: braginsky-thorne}
    \end{figure}

    We want the particles to move inertially at early and late times. Hence, we consider a setup in which all charges start very far from each other and are moving with nearly constant velocities at early times. They come together, interact, and then move away. At late times, their velocities are once again nearly constant. During the interacting phase, they emit radiation due to their accelerations. The setup is illustrated in Fig. \ref{fig: braginsky-thorne}, and it essentially consists of a general scattering problem. We can also allow the charges to collide and move away together, or for an incoming charge to split into two or more pieces. This does not affect the calculations significantly.

    At late times, the vector potential for the system comprised of all these charges will be given by
    \begin{equation}
        \vb{A}_{\text{out}}(\vb{r}) = \frac{\mu_0}{4 \pi r} \sum_{\substack{i \\ \text{out}}} \frac{Q_i \vb{v}_i}{1 - \vu{r} \vdot \vb{v}_i/c} + \order{\frac{1}{r^2}}.
    \end{equation}
    This is just Eq. \eqref{eq: vector-potential-inertial} summed over each particle. A similar expression holds at early times. Hence, the change in the vector potential is 
    \begin{equation}\label{eq: braginsky-thorne}
        \Delta\vb{A} = \frac{\mu_0}{4 \pi r} \qty[\sum_{\substack{i \\ \text{out}}} \frac{Q_i \vb{v}_i}{1 - \vu{r} \vdot \vb{v}_i/c} - \sum_{\substack{j \\ \text{in}}} \frac{Q_j \vb{v}_j}{1 - \vu{r} \vdot \vb{v}_j/c}] + \order{\frac{1}{r^2}}.
    \end{equation}
    The analog of Eq. \eqref{eq: braginsky-thorne} for gravitational waves is known as the Braginsky--Thorne formula \cite{braginsky1987GravitationalwaveBurstsMemory}.

    Using Eqs. \eqref{eq: memory-vector-potential} and \eqref{eq: braginsky-thorne}, we can compute the memory vector due to the scattering of many charges. It is given by 
    \begin{equation}\label{eq: braginsky-thorne-memory-vector}
        \vb*{\Delta} = - \frac{\mu_0}{4 \pi r} \qty[\sum_{\substack{i \\ \text{out}}} \frac{Q_i \vb{v}_i}{1 - \vu{r} \vdot \vb{v}_i/c} - \sum_{\substack{j \\ \text{in}}} \frac{Q_j \vb{v}_j}{1 - \vu{r} \vdot \vb{v}_j/c}]^{\text{T}} + \order{\frac{1}{r^2}}.
    \end{equation}
    If we know the initial and final velocities of each of the scattering charges, we can obtain the memory kick on a test particle. These velocities will generically be different in a nontrivial scattering process. Using the memory vector, we can then compute other quantities such as the large gauge transformation \(\lambda\), or the radial change in the Coulombic field to leading order \(\vu{r}\vdot\Delta \vb{E}\) (Exercise S.6 in the supplementary material). Interestingly, we can also do the converse: if we know \(\vu{r}\vdot\Delta \vb{E}\), it is possible to reconstruct \(\vb*{\Delta}\) (Exercise S.7 in the supplementary material).
    
\section{Measuring Memory}\label{sec: experimental-prospects}
    To this day, no memory effects have ever been experimentally observed. By this, we mean that it has never been experimentally verified that the memory vector (or its analogues in other theories) is nonvanishing in a certain process. How did we not notice such an effect in such a well-studied field?

    A key assumption in Eq. \eqref{eq: braginsky-thorne-memory-vector} is that the source charges move inertially at early and late times. Most often, this is not how currents and charges are handled in a laboratory. Electrons in a wire are confined to the wire, and thus their movement is bounded. Since they are only allowed to move in a finite region of space, they cannot be moving inertially and at finite speeds for arbitrarily long times. In contrast, scattering experiments involving particle beams have currents that are too small to give a visible effect~\cite{bieri2024ExperimentMeasureElectromagnetic}. 

\begin{figure}[tb]
        \centering
        \includegraphics{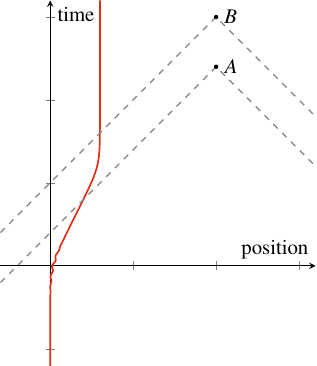}
        \caption{Minkowski diagram illustrating how memory can be measured in a ``fast'' experiment. In practical situations, the motion of the source charge (illustrated by a dark solid line) will not be inertial at a finite speed forever. For instance, the charge may hit the laboratory's wall. In this diagram, we illustrate this by making the source return to rest after a finite time, thus having the same velocity as at the beginning of the experiment and erasing any memory from the electromagnetic field. If an experimentalist makes their memory measurement sufficiently fast (for instance, at \(A\)), then the effects of the charge not being eternally in motion with a constant finite velocity can be bypassed. If the measurement is made too late (for instance, at \(B\)), then the memory will have been erased due to the source not being able to move indefinitely at a finite constant speed.}
        \label{fig: trajectory-vanishing-velocity}
    \end{figure}

    Importantly, however, the ``late'' and ``early'' times considered in Eq. \eqref{eq: braginsky-thorne-memory-vector} need not be infinite. In fact, they only give the integration range for Eq. \eqref{eq: kick-general-formula}. As long as the integral runs over the pulse of radiation, the approximate memory vector would be nonvanishing, even if the integral is computed over a short interval. We saw this in Fig. \ref{fig: Ay-oscillating}: the shift in the potentials can be seen even if ``early'' and ``late'' times are chosen to be just around the radiation pulse, but still relatively small. In this sense, we do not need the source charges to move at constant finite speeds forever. But we need to pay attention to them. More precisely, we need to make the measurements before the finite size of the laboratory interferes and brings the sources to a halt. This is illustrated in a spacetime diagram in Fig. \ref{fig: trajectory-vanishing-velocity}. In summary, memory should still be measurable during a finite-time experiment, as long as the measurement is carried out sufficiently fast \cite{bieri2024ExperimentMeasureElectromagnetic}.

    We recall that memory is measurable at finite distances. As we have shown in Sec. \ref{sec: large-gauge-transformations}, measuring at a finite distance does lead to deviations---such as the slope in the top panel of Fig. \ref{fig: Ay-oscillating}. However, this can be addressed by making the measurement sufficiently far away, or sufficiently fast. The key expression is Eq. \eqref{eq: large-r-retarded-time}, which relates the distance to the source and how much retarded time we expect to probe. 

    The Liénard--Wiechert setup is not convenient for measuring memory in a laboratory, since it would require scattering charges. Nevertheless, \textcite{bieri2024ExperimentMeasureElectromagnetic} have proposed an experimental setup based on a radiating antenna that is also capable of imparting memory on the electromagnetic field. We refer the reader to their work for further details, and to Exercise S.8 in the supplementary material for a short discussion of memory with continuous sources. 
    
    Throughout this article, we focused on charges moving in vacuum. As a consequence, we considered only sources that move slower than light. However, interesting consequences happen once we consider memory sourced by superluminal charges moving in material media. It was shown by \textcite{zosso2025EnhancementElectromagneticMemory} that memory can be significantly enhanced when the source charge's final velocity is superluminal, which could improve the prospects of an experimental measurement. We refer the reader to the original paper for more details.

\section{Conclusion}\label{sec: conclusion}
    We have discussed several features of memory in electrodynamics. Over the course of these discussions, we have answered each of the five questions we had posed at the end of Sec. \ref{sec: background}. With our newly acquired knowledge, let us revisit them and see how this happened.
    \begin{enumerate}
        \item Which physical scenario, if any, allows the integral in Eq. \eqref{eq: kick-general-formula} to be nonvanishing at order \(1/r\)?
        
        As we have seen in Sec. \ref{sec: field-lines}, the Liénard--Wiechert solution allows the integral in Eq. \eqref{eq: kick-general-formula} to be nonvanishing at order \(1/r\). There is an important caveat: the source particle needs to have a final velocity that differs from its initial velocity. This occurs naturally in scattering problems, as we pointed out in Sec. \ref{sec: radiation-from-scattering}. This is difficult to implement experimentally, since the motion of the sources is bounded by the physical size of the laboratory. As pointed out, in Sec. \ref{sec: experimental-prospects}, there are good prospects to overcoming this. 
        \item What is a vacuum in electrodynamics, and how can these different vacua be physically realized?
        
        A ``vacuum'' is a field configuration with no electromagnetic waves. As we pointed out in Secs. \ref{sec: LW-solution} and \ref{sec: field-lines}, these configurations can occur easily in the Liénard--Wiechert solution. Different vacua are obtaind by considering sources moving with different constant velocities. If a source charge first moves at some initial velocity, accelerates, and then moves at a new final velocity, the electromagnetic field will undergo a ``vacuum transition.'' This is illustrated in Fig. \ref{fig: field-lines}.
        \item Can the memory effect be measured in our universe at a finite distance, during a finite time, in a real laboratory? Or would it be necessary to carry a measurement during infinite time, or at an infinite distance, to see the effect?
        
        Yes, it can be measured in our universe at a finite distance, and during a finite time. As we discussed in Sec. \ref{sec: large-gauge-transformations}, finite-distance effects can make the effect difficult to observe. For example, Fig. \ref{fig: Ay-oscillating} shows that the characteristic ``step'' in the vector potential slowly dies off at short distances. Nevertheless, the same figure shows that larger  distances make the step persist for longer. As we pointed out in Sec. \ref{sec: experimental-prospects}, a sufficiently fast measurement should then be able to probe the memory effect. The term ``sufficiently fast'' is characterized by Eq. \eqref{eq: large-r-retarded-time}.
        \item Do very-low-energy particles (``soft photons'') really lead to physically observable effects, or are they automatically unphysical?
        
        They do lead to physically observable effects---namely, the memory effect! In Sec. \ref{sec: background} we showed that the Fourier transform of a configuration with memory is strongly dominated by low-frequency modes. Yet, as we just argued above, the memory effect should be observable. 
        \item If the memory effect is indeed physical, how is it related to gauge transformations?
        
        This relation shows up in Sec. \ref{sec: large-gauge-transformations}. If we consider the vector potential at early and late times, it undergoes a change that is, mathematically, a gauge transformation. However, this transformation is ``large,'' in the sense it does not vanish at infinity. This leads to several subtleties, such as the conservation laws that we alluded to in Sec. \ref{sec: background}.
    \end{enumerate}

    While our discussions focused on a specific aspect of infrared physics of a specific theory, it is important to point out that these investigations extend across other topics in high-energy physics and gravitational physics. The gravitational wave displacement memory effect is particularly notable.

    \begin{figure}[tb]
        \centering
        \includegraphics{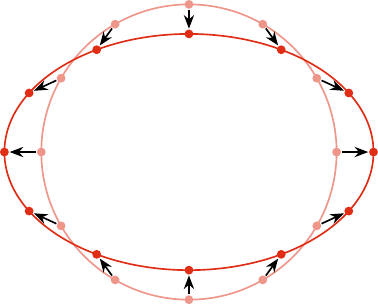}
        \caption{Gravitational waves distort a circular array of particles into an ellipse, with the polarization of the wave dictating its orientation. The gravitational wave displacement memory effect is a permanent shift in the distance between two freely falling bodies. The polarization of the gravitational wave determines if the bodies get closer or farther apart.}
        \label{fig: gw-displacement}
    \end{figure}

    In the case of gravitational waves, \textcite{zeldovich1974RadiationGravitationalWaves,braginsky1987GravitationalwaveBurstsMemory} originally imagined many stars or black holes scattering off of each other, and possibly merging. Just as in the electromagnetic scenario, the accelerations of these massive bodies would lead to the emission of gravitational radiation and a subsequent memory effect. Due to the intricacies of general relativity, the gravitational memory effect has a few different nuances. For example, it can receive important nonlinear contributions \cite{christodoulou1991NonlinearNatureGravitation}. Furthermore, instead of a velocity kick, it leads to a permanent displacement. If a gravitational wave passes through two nearby freely falling particles, then the final distance between the particles will generally be different from the initial distance. In general relativity, the final distance is sensitive to the analog of the electromagnetic potentials. Whether the particles get closer or farther away depends on the polarization of the gravitational wave, as illustrated in Fig. \ref{fig: gw-displacement}.

    As its electromagnetic counterpart, the gravitational wave displacement memory effect has not yet been observed. The main difficulty is that the current gravitational wave observatories (LIGO, Virgo, and KAGRA) are not very sensitive to low-frequency signals. It is not possible to distinguish memory from the noise measured by the detectors. Nonetheless, combining the evidence for a memory effect gathered from multiple different events can lead to a statistically significant result; a few thousand binary black hole mergers would be needed to find definitive evidence for gravitational wave memory. Moreover, future gravitational wave observatories are expected to be able to measure the memory effect in some individual events. For further discussions on measuring gravitational wave memory, see, for example, Refs. \onlinecite{grant2023OutlookDetectingGravitationalwave,*grant2023ErratumOutlookDetecting,zosso2026TowardClaimingDetection} and the references therein.

    On the theoretical side, the gravitational wave displacement memory effect also abides by an infrared triangle such as the one in Fig. \ref{fig: infrared-triangle}. The analogs of large gauge transformations are known as `supertranslations,' and there is also a soft graviton theorem---which is also due to \textcite{weinberg1965InfraredPhotonsGravitons}. See, for example, Refs. \onlinecite{aguiaralves2025LecturesBondiMetzner,strominger2018LecturesInfraredStructure} for introductions.

    The memory effect is a magnificent reminder of how much humankind still has to learn about the universe. Even in theories as established and understood as electrodynamics, there may still be unobserved phenomena patiently waiting to be discovered. We hope to have convinced the reader that some of these can be investigated starting from textbook physics.

\section*{Supplementary Material}

    This paper is accompanied by supplementary material, which includes suggested exercises, their solutions, the \textsc{Asymptote} \cite{bowman2008AsymptoteVectorGraphics,hammerlindl2004AsymptoteDescriptiveVector} code used to produce Fig. \ref{fig: field-lines}, and \textsc{Mathematica} \cite{wolframresearch2026Mathematica150} code that can be used to reproduce Figs. \ref{fig: step-function} and \ref{fig: Ay-oscillating}. This material is available through the arXiv version's ancillary files.

\begin{acknowledgments}
    Part of the calculations in this work were carried out with the aid of \textsc{Mathematica 15.0} \cite{wolframresearch2026Mathematica150}. Figure \ref{fig: field-lines} was drawn using \textsc{Asymptote} \cite{bowman2008AsymptoteVectorGraphics,hammerlindl2004AsymptoteDescriptiveVector}.

    The work of NAA was supported by the São Paulo Research Foundation (FAPESP) under process 2025/05161-0.
\end{acknowledgments}

\section*{Author Declarations}
    \subsection*{Author Contributions}
        Both authors contributed equally to this work and share first authorship.

\bibliography{bibliography}
\end{document}